\documentclass[prl,aps,twocolumn,showpacs,superscriptaddress]{revtex4}
\pdfoutput=1

\usepackage{amsmath,amssymb,graphicx,epstopdf,eps fig,subfigure,color}

\begin{document}

\title{\bf Exploiting quantum coherence of polaritons for ultra sensitive detectors}
\author {Guido Franchetti}
\affiliation {Department of Applied Mathematics and Theoretical Physics,
University of Cambridge,  Cambridge, CB3 0WA}
\author{Natalia G. Berloff}
\affiliation {Department of Applied Mathematics and Theoretical Physics,
University of Cambridge,  Cambridge, CB3 0WA}
\author{Jeremy J. Baumberg}
\affiliation{NanoPhotonics Centre, Cavendish Laboratory, University of Cambridge, Cambridge, CB3 0HE}
\date{\today}

\begin {abstract} Besides being superfluids, microcavity exciton-polariton condensates are capable of spontaneous pattern formation due to their forced-dissipative dynamics. Their macroscopic and easily detectable response to small perturbations can be exploited to create sensitive devices. We show how controlled pumping in the presence  of a peak-dip shaped potential can be used to detect small externally applied velocities which lead to the formation of traveling holes in one dimension and vortex pairs in two dimensions.
Combining an annulus geometry with a weak link, the set up that we describe can be used  to create a sensitive polariton gyroscope.
\end{abstract}
\maketitle
Microcavity exciton-polaritons are two-dimensional half-light half-matter quasi-particles
that result from the hybridisation of quantum well excitons
and photons in a planar Fabry-Perot resonator. At low enough densities, they behave as
bosons and if the temperature is lower than some critical value they undergo Bose-Einstein condensation \cite{Balili:2007gc,kasprzak06:nature}. Recent experiments have investigated exciton-polariton condensation and the phenomena associated with it, such as pattern formation \cite{Tosi:2012ik,Manni:2011ku},
quantised vortices and solitons \cite{Lagoudakis:2008ia,Amo:2011bf}, increased coherence and the cross-over to regular lasing; recent reviews of the field can be found in \cite{Keeling:2011ho,Carusotto:2012vzxx}.
These discoveries have opened a path for new technological developments in optoelectronics (ultra-fast optical switches, quantum circuits), medicine (compact terahertz lasers) and power production
(hybrid organic-inorganic solar cells).
However one fundamental property of exciton-polariton condensates, their ability to flow without friction at moderate velocities \cite{Amo:2009bl,Wouters:2010ee},
 has never been considered for technological use. The existence of frictionless flow in polariton condensates sets them apart
from other solid-state quantum systems and links them to cooperative fluids, such as superfluid helium
or atomic condensates. In this letter we  propose an idea 
that may lead to a new generation of ultra-sensitive devices based on the  quantum coherence and superfluidity of exciton-polariton
condensates.
The important ingredients of this novel technology come from the unique
properties of polariton condensates: (1) they are non-equilibrium systems capable of pattern
formation; (2) their dynamics is controlled by the balance between gain, due to
continuous pumping, losses and non-linearity; (3) polaritons condense at relatively high (even room) temperature thanks to their very
small effective mass; 
 (4) one can easily engineer any external landscape and vary pumping in space and
time;
 (5) polariton condensates form quantised vortices in response to slight changes in
the environment: when flow exceeds the critical velocity, when
fluxes interact, when pumping powers exceed a threshold for pattern forming instabilities,
when the magnetic field exceeds a threshold, etc. 
These properties allow one to prepare the system in a state slightly below the criticality for vortex formation, so that a tiny external
perturbation will take it over the criticality, leading to a macroscopic and easily
detectable response. Similar principles underpin superconducting single photon detectors, biased just below their transition temperature. Therefore, we propose  exploiting  superfluid properties and sensitivity
to vortex formation of polariton condensates to create sensitive devices that respond to
slight changes in rotation rate, gravity or magnetic field. We will illustrate the main principles
using the example of an exciton-polariton gyroscope.
 
Superfluid helium gyroscopes have already been shown to provide sensitive means for detecting absolute
rotations \cite{Avenel:1997io,Packard:1992jz}, but required very low temperature for  operation. 
A superfluid 
\begin{figure}[htbp]
 \centering
\bigskip
\vskip -1.7em
\includegraphics[height=2 in]{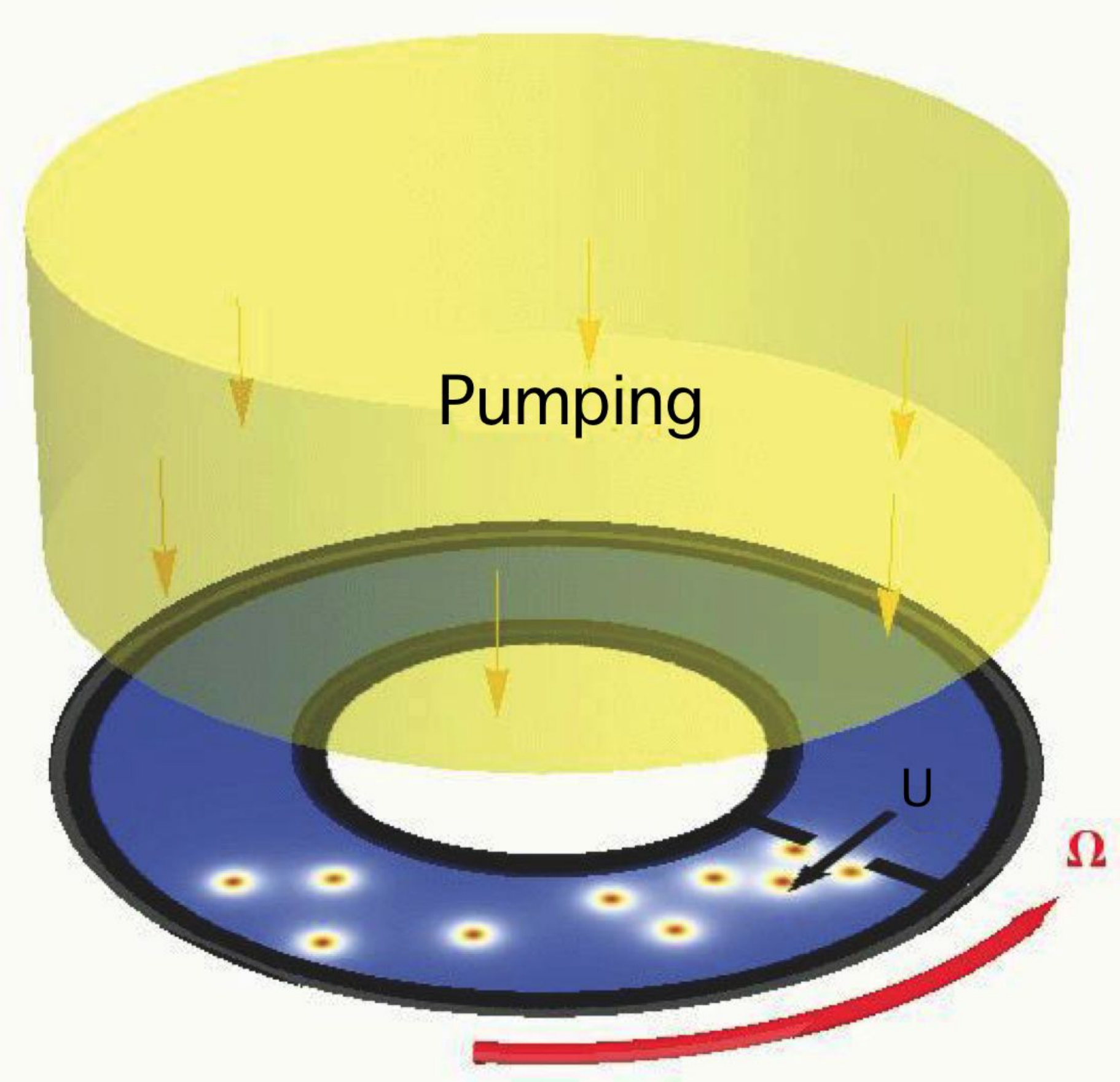}
\caption{ (color online) Schematic of the experiment: pumped polariton condensate channel with a weak link. If the flow across the link is high enough, vortices are generated.}
\vskip -0.7em
\label{gyro}
\end{figure}
in an annulus (ring) that is slowly rotated  remains motionless as
there is no friction with boundaries.
But if a partition with a small opening (weak link) is inserted in the annulus, a flow in the direction opposite to the rotation will be generated through it. To leading order \cite{Packard:1992jz}, the velocity $U$ across the opening is related to the angular speed of rotation $\Omega$ by the relation $U =\Omega R^2/\delta $, where $R$ ($\delta$) is the width of the annulus (opening). When $U$ exceeds some critical value, which is of the order of the Landau critical velocity $U_c$,  vortices are generated and detected as phase slips. For the case of superfluid He${}_4$, $U_c \sim 50\, m/s, R \sim 0.1\,m, \delta \sim 100\,nm$, allowing accurate detection of the Earth's daily rotation rate. It was also established that superfluid rotation sensors,
similar to  atomic beam gyroscopes, belong to the
same class of quantum interference effects as  Sagnac
light-wave experiments \cite{Varoquaux:2008}. The same principles can be applied to a polariton gyroscope, schematic of which is given in Fig.~\ref{gyro}, which offers the advantage of potentially operating at room temperatures. The  critical velocity  for  vortex formation in polariton condensates \cite{Wouters:2010ee}  is three-four orders of magnitude higher  than that in  superfluid He${}_4$, a fact that would reduce the sensitivity of the polariton gyroscope. However, the pattern forming properties of polariton condensates allow one to severely reduce the velocities needed for the vortex formation to occur and to use  pumping intensity as the control parameter.
We propose to insert a peak-dip shaped potential, which will further accelerate the flow, at the position of the weak link.
Such a potential can be prepared either using a combination of etching \cite{Dasbach:2001wt,Kaitouni:2006hr,Bajoni:2008ef} and stress induced traps \cite{Balili:2007gc}, or directly defining blue-shifted trap potentials via spatial light modulators \cite{Tosi:2012ik}.
Even in the absence of any externally applied flow, velocity fluxes connecting regions where the density is low (at the potential peak) to regions where the density is high (at the potential dip) will be generated; in between them there will be a point where the condensate speed takes its maximum value. Close to such a point, the density has a local minimum (Bernoulli effect).
A slight change of external conditions, e.g. an increase of the potential strength, or a decrease of the pumping intensity, will  further lower the density. If the perturbation is large enough, the density minimum will reach zero and a vortex pair will be emitted, as we   illustrate below using a mean-field model of polariton condensates.
The scheme for measuring rotations then becomes quite
straightforward: 
having prepared a potential such that the system is in a slightly subcritical configuration for the range of rotation speeds that one intends to measure,  the strength of the pumping intensity is varied, recording the moment
when vortices start to nucleate.  From this one deduces the back-flow through the
aperture and, therefore, the rotational velocity.

Polariton condensates at temperatures much lower than the critical temperature for condensation can be effectively 
described using the complex Ginzburg-Landau equation (cGL) \cite{Keeling:2008hj,wouters07:goldstone}. If a condensate is flowing with bulk speed $U$ along the $x$ axis, the cGL  in a frame co-moving with the condensate  takes the form
\begin{equation}
  \label{eq:cgpe}
  \begin{split}
  2  (\eta \! -\! i )\!\left( \partial _t\!   +\!  U \partial _x   \right)  \psi  \!=\!  
  \!\Big[\! \nabla ^2 \!-\!    V(x,y) \!-\!| \psi | ^2 
\!-\! i \Big(\! \alpha\!   -\! \sigma |\psi| ^2  \!\Big) \! \Big ] \psi  
  \end{split}
  \end{equation}
where $V(x,y)$ describes an external potential, $\alpha$ is
a  pump rate,  $\sigma$ is a nonlinearity
which causes pumping to reduce as density increases and $\eta \sim 0.1$ is an energy relaxation term \cite{Wouters:2010vyxx} which causes the evolution towards  states of lower energy. Linear losses are included in $\alpha$.
Eq.~(\ref{eq:cgpe}) is stated in harmonic oscillator units, assuming that $V=\omega r^2/2$ and measuring 
energy in units of 
$\hbar\omega/2$, length in units of the oscillator length
$l=\sqrt{\hbar/m\omega}$, and time in units of $\omega^{-1}$, where
$\omega$ is the oscillator frequency of the trapping potential.
For $m$ and $\hbar\omega$  we take the values found  by Balili et
al \cite{Balili:2006dm,Balili:2007gc}: 
$\hbar\omega=0.066$~meV, 
$m=7\times10^{-5}m_e$
and for $\alpha$ and $\sigma$ we follow estimates given in
 Refs.~\cite{Keeling:2008hj,Borgh:2010jv}:
 $\sigma =0.3 $, $0< \alpha < 10$.

\begin{figure}[th!]
\begin{center}
\includegraphics[width=0.48\textwidth]{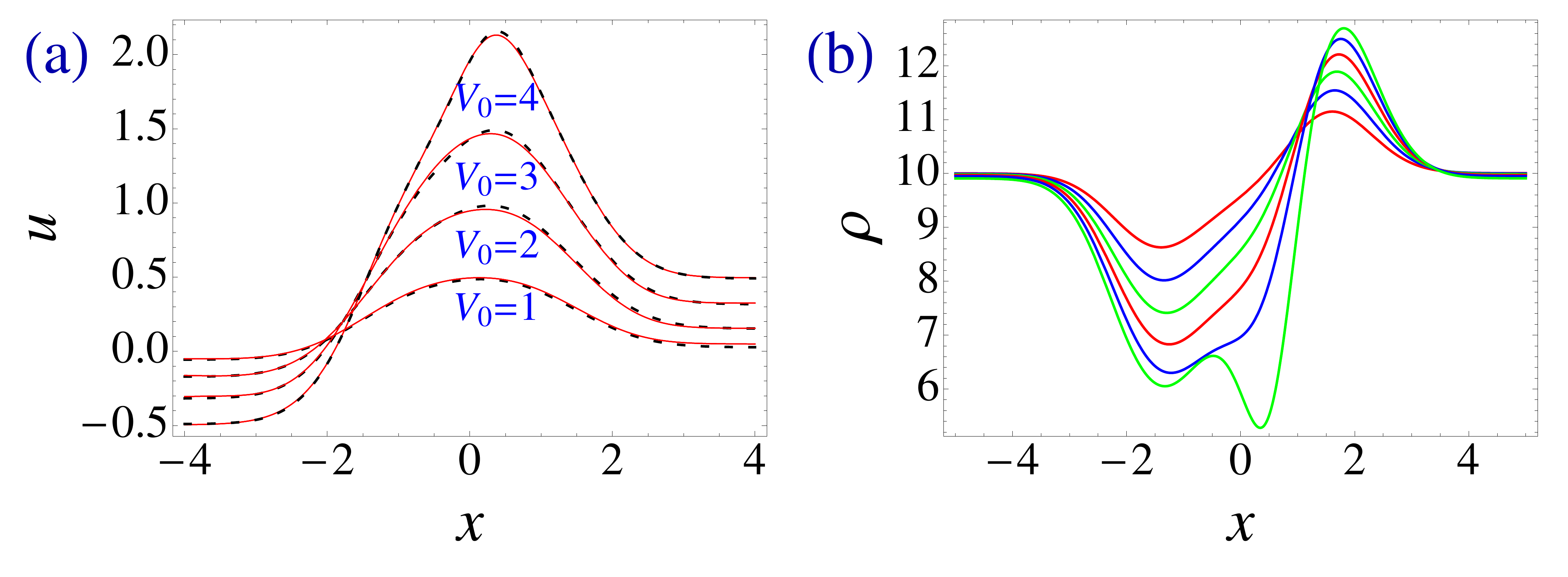}
\caption{(color online) (a) Numerical (black dashed line) and analytical estimates (red solid line) of velocity profiles as $V _0 $ is increased from $1$ to $4$. (b) Numerical solutions for $\rho$ with $ V _0=1.4, 2.0, 2.6,  3.2, 3.8, 4.4.$ Higher values of $V _0 $ correspond to lower values of  min($\rho$).  In both figures $x _0 =1.5 $.}
\label{numtheo}
\end{center}
\vskip -2em
\end{figure}

\begin{figure}[h!]
\begin{center}
\includegraphics[width=0.45\textwidth]{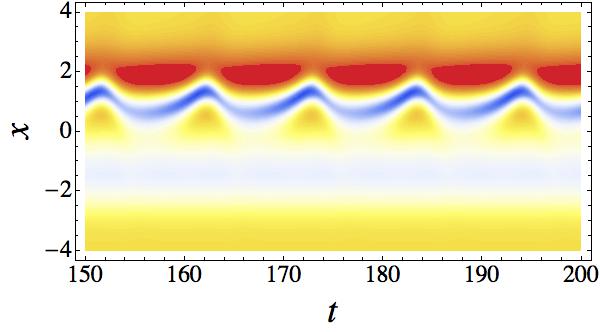}
\vskip -1em
\caption{(color online) Traveling holes, identified by the localized low-density blue region, form around $x =0 $, and travel up to $x \sim 1.5 $. The process repeats periodically in time. ($V _0 =4.8$, $x _0 =1.5 $, $\alpha =4$, $U =0$).}
\label{gsol}
\end{center}
\vskip -2.3em
\end{figure}

In order to illustrate the main principles of the vortex formation mechanism, we consider first the simpler 1D case, with an external potential of the form
$\tilde V (x) = V _0 \left( \exp \left( -(x-x _0 )^2\right)  - \exp (- (x + x _0 )^2 ) \right) $.  The peak-dip shape of this potential allows for a better acceleration of the condensate than a single bump; moreover, in 2D the dip has also the useful role of  temporarily trapping vortices and making  their detection easier. After performing a Madelung transformation setting $\psi =\sqrt{ \rho } \exp (i \phi ) $, equation (\ref{eq:cgpe}) in a frame at rest ($U =0 $) and for a stationary state becomes the system
\begin{align}
\label{mad} 
&\partial _x ( \rho u ) =\left(  \alpha - \sigma \rho - \eta \mu \right) \rho \\
\label{mad2} 
&\mu =\rho + u ^2  + V  (x) - \partial  ^2 _x \sqrt{ \rho }/ \sqrt{ \rho }
\end{align}
where $\mu$ is the chemical potential. 
The speed $u =\nabla \phi $ takes its maximum at some point  between $- x _0  $ and $x _0 $, the maximum and minimum of the potential.
It follows from Eq.~(\ref{mad2}), which is a generalized version of Bernoulli's equation, that close to the maximum of the speed, density has a local minimum.
The numerical results displayed in Fig.~\ref{numtheo} show that max($u$) is an increasing function of $V _0  $ and  min($ \rho  $)  a decreasing one.  This can be  verified analytically deriving an approximate solution  as follows. For very small $V _0 $ it is natural to expect a dip-peak density profile; 
however, for higher values of $V _0 $, there will be an additional minimum due to the Bernoulli effect.
 Therefore we take a density ansatz of the form 
$\rho = $\mbox{$\alpha / (\sigma + \eta ) \!+\! B \exp(-b(x-x _3 ) ^2 )\!+\!C \exp(-c(x-x _1 ) ^2 )$} $-Q \exp(-q(x-x _2 ) ^2 )$, where $b,B,c,C,q,Q,x_1,x_2,x_3$ are free parameters.
Substituting in  (\ref{mad}) and integrating one has:
\begin{align*}
& \rho u \!= \!
 \frac{\sqrt{ \pi }\alpha \sigma }{2 (\sigma \!+\! \eta ) }\!\Bigg[\!\!
	\frac{Q }{\sqrt{ q }}  \text{erf}\left(\!\sqrt{q} \tilde x_3\! \right)\!
	-\!\frac{B }{\sqrt{ b }}\text{erf} (\!\sqrt{b} \tilde x _1 )\!
	-\!\frac{C }{\sqrt{ c }} \text{erf}\left(\!\sqrt{c} \tilde x _2 \!\right)
\!\!\Bigg]
\\ & 
-\! \sqrt{ \frac{\pi }{2} } \!\frac{\sigma }{2} \Bigg [ 
	\frac{B^2}{ \sqrt{b}}  \text{erf}(\!\sqrt{2b} \tilde x _1 ) \!+\! 
	\frac{C^2}{ \sqrt{  c }}  \text{erf}(\!\sqrt{2c} \tilde x _2 ) \!+ \!
	\frac{ Q^2 }{\sqrt{ q }} \text{erf}(\!\sqrt{2q} \tilde x_3 )\!
\Bigg ]
\\& + \sqrt{ \pi } \sigma \Bigg[ 
-E(B,C,b,c,x _1 , x _2 ) + E(B,Q,b,q,x _1, x_3 ) \\ &\phantom{ + \sqrt{ \pi } \sigma \Bigg[ }+ E(C,Q,c,q,x_2, x _3)
   \Bigg ]
   \end{align*} 
where $\tilde x _i \equiv x -x _i $, $x _{ ij } \equiv x _i - x _j $ and 
\begin{equation*}
E(B,C,b,c,x _i , x _j )\!=\!\! \frac{BC}{\sqrt{ b + c }}\, \mathrm{e}^ {-\left(\frac{b\,c\, x _{ ij }^2}{b + c }    \right)}  \mathrm{erf } \!\left(  \frac{b \tilde x _i +c \tilde x _j }{\sqrt{ b +  c }}\right)  
\end{equation*}
The potential $V (x) $ is then determined from (\ref{mad2}) and contains the unknown parameters $B, C, Q, b, c, q, x _1, x _2, x _3   $.
The values of these parameters can be fixed requiring $V (x) $ to fit the original potential $ \tilde V (x) $. A comparison of the analytical and numerical solutions for $u$ is shown in  panel (a) of Fig.~\ref{numtheo}, while numerical solutions for $\rho$ are shown in  panel (b).
\begin{figure}[htbp]
\begin{center}
\includegraphics[width=0.48\textwidth]{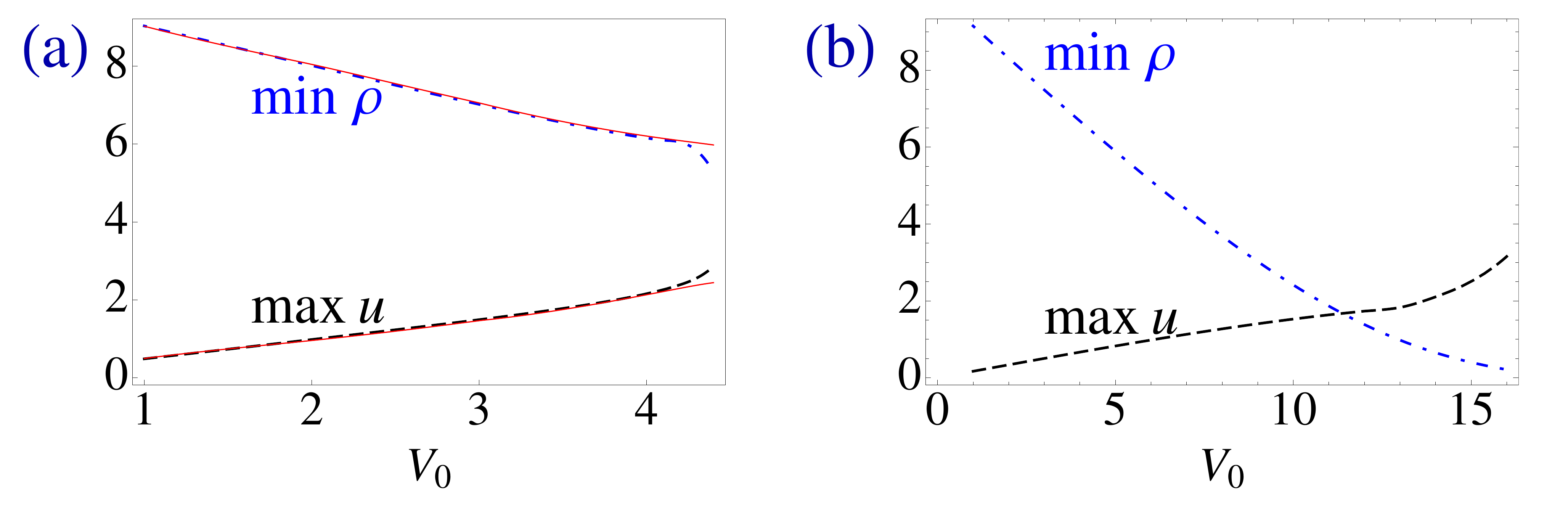}
\caption{(color online) (a) 1D. Profiles of max($u$) (numerical: black dashed; analytical: red continuous line) and of  min($\rho$)  (numerical: blue dot-dashed; analytical: red continuous line) as a function of $V _0 $, with $x _0 =1.5 $.  The theoretical solution is only valid up to $V _0  \sim 4.0 $. (b) 2D. Numerical profiles of  min($\rho(x,0)$) (blue dot-dashed line) and of  max($u _x (x,0) $) (black dashed line) as a function of $V _0 $,  $x _0 =1.5 $.}
\label{maxu}
\end{center}
\vskip -1em
\end{figure}
The behavior of max($u$)  and min($\rho$) with $V _0 $ is mostly linear up to $V _0 =4$, see Fig.~\ref{maxu} panel (a). A deviation from the linear regime leads quickly,  for $V _0 >4.4$, to the loss of stability of stationary solutions. The time dependent solutions found for $V _0 >4.4 $ are characterized by the periodic emission of traveling holes \cite{Lega:1997}, which  travel for a short time before dissipating, see Fig.~\ref{gsol}. The density hits zero when the hole is emitted.

\begin{figure}[h!]
\begin{center}
\includegraphics[width=0.45\textwidth]{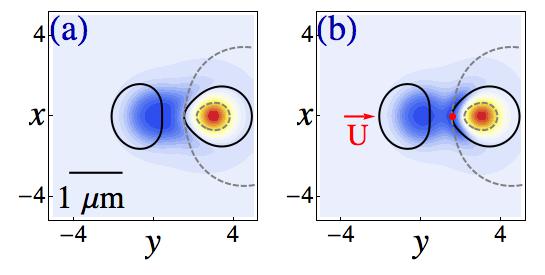}
\raisebox{22.6pt}{\includegraphics[width=0.02\textwidth,height=3.1cm]{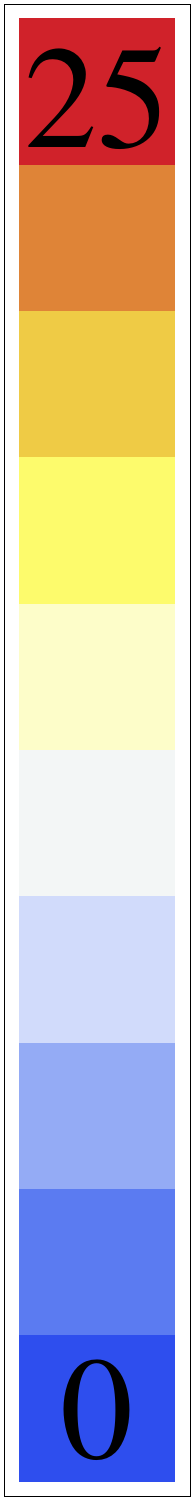}}
\vskip -1em
\caption{(color online) Real space plots of the zero level of real (black continuous line) and imaginary (grey dashed line) part of $\psi$ superimposed over the contour plot of $\rho$.   Regions of higher density are red.  (a) A slightly subcritical configuration ($\alpha =4$, $V _0  =16.8$, $x _0 =1.5 $, $U  =0 $); (b) A slightly overcritical configuration ($ \alpha =4$, $V _0  =16.8$, $x _0 =1.5 $, $U =0.2 $) at the moment of vortex pair formation. A red arrow shows the direction of the flow; a red dot shows the point where the vortex pair is being generated.
}
\label{reimgrid}
\end{center}
\vskip -2em
\end{figure}

\begin{figure}[htbp]
\begin{center}
\includegraphics[width=0.45\textwidth]{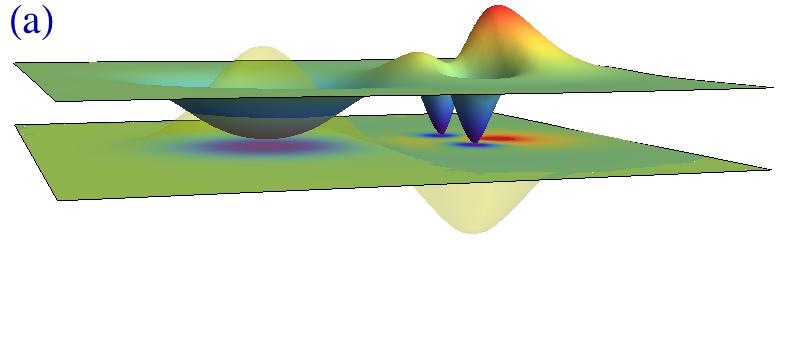}
\vskip -3em
\includegraphics[width=0.45\textwidth]{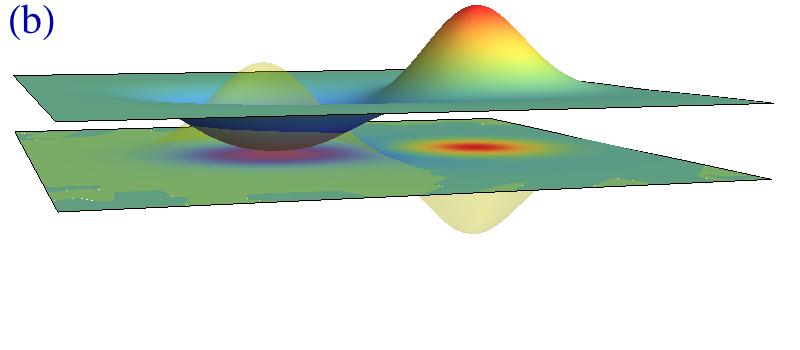}
\vskip -3em
\caption{(color online). (a) Density profile above criticality ($V _0 =14$,  $x _0 =1.5 $, $\alpha =4$, $U =0.4$). The vortex pair has been temporarily trapped by the dip in the potential.  (b) Density profile for a stationary solution below criticality ($V _0 =14$, $x _0 =1.5 $, $\alpha =4$, $U   =0$). The 3D plots show the density profile, with the potential superimposed in yellow. The 2D contour plot gives the projection of the density values on the $x y $ plane, with regions of higher density colored in red. The direction of the flow is from the left to the right.}
\label{vortplot}
\end{center}
\vskip -1em
\end{figure}

In two dimensions we take a similar potential:
$ \mbox{$V\! =\!V _0\big(\! \exp(-((x+x_0)^2\!+\!y^2))\!-\exp(-((x-x_0)^2\!+\!y^2)) \big )$}$. The dynamics is similar to that of the 1D case:  min($\rho$) is again a decreasing function of $V _0 $, with a linear behavior for small values of $V _0 $, see Fig.~\ref{maxu} panel (b). When $V _0 $ exceeds a critical value, stationary solutions become unstable and vortex pairs form. If the system is close enough to criticality, the drop of min$(\rho )$ to zero and consequent onset of vortex nucleation can also be started by a small increase of the external velocity $U$ or decrease of the pumping strength $\alpha$. The process is shown in Fig.~\ref{reimgrid}, where the onset of vortex nucleation is due to an increase of  $U$. Representatives of  critical and subcritical states are shown in Fig.~\ref{vortplot}.
There is, however, one notable difference with the 1D case: while in the latter the emission of traveling holes sets in  at a finite value of min$(\rho)$, in 2D there are stationary solutions all the way down to min($\rho$)$=0 $, see the right panel of Fig.~\ref{maxu}. 
\begin{figure}[htbp]
\begin{center}
\includegraphics[width=0.45\textwidth]{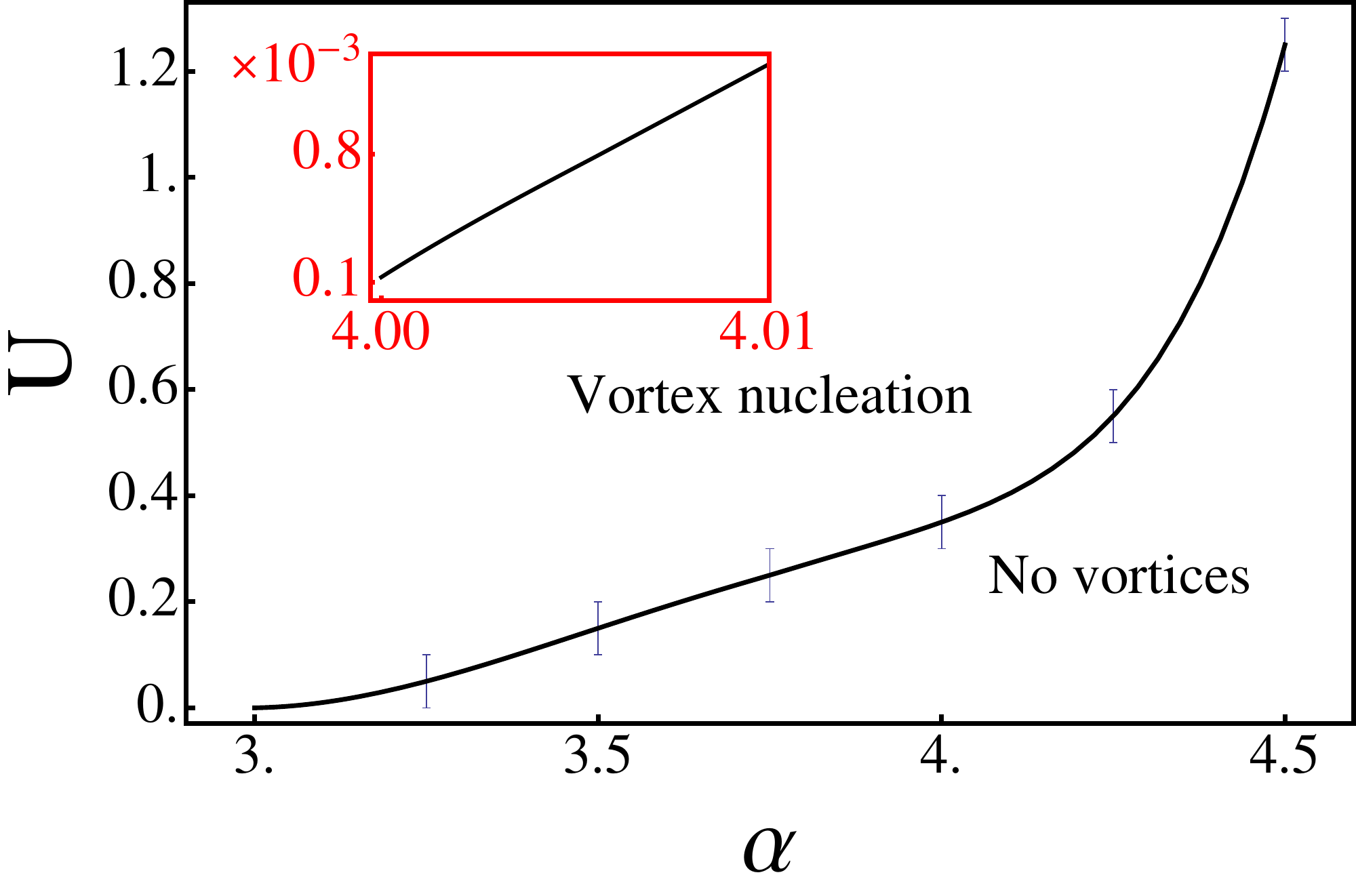}
\vskip -1em
\caption{The main plot shows part of the boundary between critical and subcritical regions of phase space with respect to the parameters $U$ and $\alpha$ for $V _0 =14$, $x _0 =1.5 $. The inset has been obtained with a different values of the potential, $V _0 =16.8 $, $x _0 =1.5 $ and gives an idea of the attainable resolution for small values of $U$.  }
\label{2dphaseplot}
\end{center}
\vskip -2em
\end{figure}
An example  calibration curve is shown in Fig.~\ref{2dphaseplot}, where the critical values of $\alpha$ which start vortex nucleation are recorded for fixed $V _0 $ and different values of $U$. The boundary  $U (\alpha) $  between subcritical and critical behavior is, for a bulk speed $U<0.5$, nearly linear, hence the sensitivity of the apparatus scales linearly with $\alpha$ or, equivalently, with the bulk density $\alpha / (\sigma  + \eta )$ of the condensate.
In the numerical simulations that we performed, the system showed a clear transition between subcritical and critical states at velocities of the order of $10^{-4}$ of the critical velocity $U_c=\sqrt{\mu/m}$ for this system.
This puts the sensitivity of polariton gyroscopes in the same range as that of superfluid helium gyroscopes, but at the advantage of potentially operating at room temperature.

In summary, we proposed an idea for creating polaritonic sensitive devices, such as a superfluid gyroscopes, based on their macroscopic response to small   perturbations. We studied the mechanism of the formation of traveling holes in 1D and vortex pairs in 2D in an externally imposed peak-dip shaped potential when the height of this potential (the difference between maximum and minimum) exceeds the threshold.
 
\begin{acknowledgments}
G.F. acknowledges funding from Marie Curie Actions ESR grants. All authors acknowledge funding from EU CLERMONT4 235114.
\end{acknowledgments}

\bibliography{biblio,polaritons}

\newcommand\textdot{\.}
\begin{thebibliography}{24}
\expandafter\ifx\csname natexlab\endcsname\relax\def\natexlab#1{#1}\fi
\expandafter\ifx\csname bibnamefont\endcsname\relax
  \def\bibnamefont#1{#1}\fi
\expandafter\ifx\csname bibfnamefont\endcsname\relax
  \def\bibfnamefont#1{#1}\fi
\expandafter\ifx\csname citenamefont\endcsname\relax
  \def\citenamefont#1{#1}\fi
\expandafter\ifx\csname url\endcsname\relax
  \def\url#1{\texttt{#1}}\fi
\expandafter\ifx\csname urlprefix\endcsname\relax\def\urlprefix{URL }\fi
\providecommand{\bibinfo}[2]{#2}
\providecommand{\eprint}[2][]{\url{#2}}

\bibitem[{\citenamefont{Balili et~al.}(2007)\citenamefont{Balili, Hartwell,
  Snoke, Pfeiffer, and West}}]{Balili:2007gc}
\bibinfo{author}{\bibfnamefont{R.~B.} \bibnamefont{Balili}},
  \bibinfo{author}{\bibfnamefont{V.}~\bibnamefont{Hartwell}},
  \bibinfo{author}{\bibfnamefont{D.~W.} \bibnamefont{Snoke}},
  \bibinfo{author}{\bibfnamefont{L.}~\bibnamefont{Pfeiffer}}, \bibnamefont{and}
  \bibinfo{author}{\bibfnamefont{K.}~\bibnamefont{West}},
  \bibinfo{journal}{Science} \textbf{\bibinfo{volume}{316}},
  \bibinfo{pages}{1007} (\bibinfo{year}{2007}).

\bibitem[{\citenamefont{Kasprzak et~al.}(2006)\citenamefont{Kasprzak, Richard,
  Kundermann, Baas, Jeambrun, Keeling, Marchetti, Szymanska, Andre, Staehli
  et~al.}}]{kasprzak06:nature}
\bibinfo{author}{\bibfnamefont{J.}~\bibnamefont{Kasprzak}},
  \bibinfo{author}{\bibfnamefont{M.}~\bibnamefont{Richard}},
  \bibinfo{author}{\bibfnamefont{S.}~\bibnamefont{Kundermann}},
  \bibinfo{author}{\bibfnamefont{A.}~\bibnamefont{Baas}},
  \bibinfo{author}{\bibfnamefont{P.}~\bibnamefont{Jeambrun}},
  \bibinfo{author}{\bibfnamefont{J.~M.~J.} \bibnamefont{Keeling}},
  \bibinfo{author}{\bibfnamefont{F.~M.} \bibnamefont{Marchetti}},
  \bibinfo{author}{\bibfnamefont{M.~H.} \bibnamefont{Szymanska}},
  \bibinfo{author}{\bibfnamefont{R.}~\bibnamefont{Andre}},
  \bibinfo{author}{\bibfnamefont{J.~L.} \bibnamefont{Staehli}},
  \bibnamefont{et~al.}, \bibinfo{journal}{Nature}
  \textbf{\bibinfo{volume}{443}}, \bibinfo{pages}{409} (\bibinfo{year}{2006}).

\bibitem[{\citenamefont{Tosi et~al.}(2012)\citenamefont{Tosi, Christmann,
  Berloff, Tsotsis, Gao, Hatzopoulos, Savvidis, and Baumberg}}]{Tosi:2012ik}
\bibinfo{author}{\bibfnamefont{G.}~\bibnamefont{Tosi}},
  \bibinfo{author}{\bibfnamefont{G.}~\bibnamefont{Christmann}},
  \bibinfo{author}{\bibfnamefont{N.~G.} \bibnamefont{Berloff}},
  \bibinfo{author}{\bibfnamefont{P.}~\bibnamefont{Tsotsis}},
  \bibinfo{author}{\bibfnamefont{T.}~\bibnamefont{Gao}},
  \bibinfo{author}{\bibfnamefont{Z.}~\bibnamefont{Hatzopoulos}},
  \bibinfo{author}{\bibfnamefont{P.~G.} \bibnamefont{Savvidis}},
  \bibnamefont{and} \bibinfo{author}{\bibfnamefont{J.~J.}
  \bibnamefont{Baumberg}}, \bibinfo{journal}{Nature Physics}
  \textbf{\bibinfo{volume}{8}}, \bibinfo{pages}{190} (\bibinfo{year}{2012}).

\bibitem[{\citenamefont{Manni et~al.}(2011)\citenamefont{Manni, Lagoudakis,
  Liew, Andr{\'e}, and Deveaud-Pl{\'e}dran}}]{Manni:2011ku}
\bibinfo{author}{\bibfnamefont{F.}~\bibnamefont{Manni}},
  \bibinfo{author}{\bibfnamefont{K.}~\bibnamefont{Lagoudakis}},
  \bibinfo{author}{\bibfnamefont{T.~C.~H.} \bibnamefont{Liew}},
  \bibinfo{author}{\bibfnamefont{R.}~\bibnamefont{Andr{\'e}}},
  \bibnamefont{and}
  \bibinfo{author}{\bibfnamefont{B.}~\bibnamefont{Deveaud-Pl{\'e}dran}},
  \bibinfo{journal}{Physical Review Letters} \textbf{\bibinfo{volume}{107}},
  \bibinfo{pages}{106401} (\bibinfo{year}{2011}).

\bibitem[{\citenamefont{Lagoudakis et~al.}(2008)\citenamefont{Lagoudakis,
  Wouters, Richard, Baas, Carusotto, Andr{\'e}, Dang, and
  Deveaud-Pl{\'e}dran}}]{Lagoudakis:2008ia}
\bibinfo{author}{\bibfnamefont{K.~G.} \bibnamefont{Lagoudakis}},
  \bibinfo{author}{\bibfnamefont{M.}~\bibnamefont{Wouters}},
  \bibinfo{author}{\bibfnamefont{M.}~\bibnamefont{Richard}},
  \bibinfo{author}{\bibfnamefont{A.}~\bibnamefont{Baas}},
  \bibinfo{author}{\bibfnamefont{I.}~\bibnamefont{Carusotto}},
  \bibinfo{author}{\bibfnamefont{R.}~\bibnamefont{Andr{\'e}}},
  \bibinfo{author}{\bibfnamefont{L.~S.} \bibnamefont{Dang}}, \bibnamefont{and}
  \bibinfo{author}{\bibfnamefont{B.}~\bibnamefont{Deveaud-Pl{\'e}dran}},
  \bibinfo{journal}{Nature Physics} \textbf{\bibinfo{volume}{4}},
  \bibinfo{pages}{706} (\bibinfo{year}{2008}).

\bibitem[{\citenamefont{Amo et~al.}(2011)\citenamefont{Amo, Pigeon, Sanvitto,
  Sala, Hivet, Carusotto, Pisanello, Lemenager, Houdre, Giacobino
  et~al.}}]{Amo:2011bf}
\bibinfo{author}{\bibfnamefont{A.}~\bibnamefont{Amo}},
  \bibinfo{author}{\bibfnamefont{S.}~\bibnamefont{Pigeon}},
  \bibinfo{author}{\bibfnamefont{D.}~\bibnamefont{Sanvitto}},
  \bibinfo{author}{\bibfnamefont{V.~G.} \bibnamefont{Sala}},
  \bibinfo{author}{\bibfnamefont{R.}~\bibnamefont{Hivet}},
  \bibinfo{author}{\bibfnamefont{I.}~\bibnamefont{Carusotto}},
  \bibinfo{author}{\bibfnamefont{F.}~\bibnamefont{Pisanello}},
  \bibinfo{author}{\bibfnamefont{G.}~\bibnamefont{Lemenager}},
  \bibinfo{author}{\bibfnamefont{R.}~\bibnamefont{Houdre}},
  \bibinfo{author}{\bibfnamefont{E.}~\bibnamefont{Giacobino}},
  \bibnamefont{et~al.}, \bibinfo{journal}{Science}
  \textbf{\bibinfo{volume}{332}}, \bibinfo{pages}{1167} (\bibinfo{year}{2011}).

\bibitem[{\citenamefont{Keeling and Berloff}(2011)}]{Keeling:2011ho}
\bibinfo{author}{\bibfnamefont{J.~M.~J.} \bibnamefont{Keeling}}
  \bibnamefont{and} \bibinfo{author}{\bibfnamefont{N.~G.}
  \bibnamefont{Berloff}}, \bibinfo{journal}{Contemporary Physics}
  \textbf{\bibinfo{volume}{52}}, \bibinfo{pages}{131} (\bibinfo{year}{2011}).

\bibitem[{\citenamefont{Carusotto and Ciuti}(2012)}]{Carusotto:2012vzxx}
\bibinfo{author}{\bibfnamefont{I.}~\bibnamefont{Carusotto}} \bibnamefont{and}
  \bibinfo{author}{\bibfnamefont{C.}~\bibnamefont{Ciuti}}
  (\bibinfo{year}{2012}), \eprint{arXiv:1205.6500v1}.

\bibitem[{\citenamefont{Amo et~al.}(2009)\citenamefont{Amo, Lefr{\`e}re,
  Pigeon, Adrados, Ciuti, Carusotto, Houdre, Giacobino, and
  Bramati}}]{Amo:2009bl}
\bibinfo{author}{\bibfnamefont{A.}~\bibnamefont{Amo}},
  \bibinfo{author}{\bibfnamefont{J.}~\bibnamefont{Lefr{\`e}re}},
  \bibinfo{author}{\bibfnamefont{S.}~\bibnamefont{Pigeon}},
  \bibinfo{author}{\bibfnamefont{C.}~\bibnamefont{Adrados}},
  \bibinfo{author}{\bibfnamefont{C.}~\bibnamefont{Ciuti}},
  \bibinfo{author}{\bibfnamefont{I.}~\bibnamefont{Carusotto}},
  \bibinfo{author}{\bibfnamefont{R.}~\bibnamefont{Houdre}},
  \bibinfo{author}{\bibfnamefont{E.}~\bibnamefont{Giacobino}},
  \bibnamefont{and} \bibinfo{author}{\bibfnamefont{A.}~\bibnamefont{Bramati}},
  \bibinfo{journal}{Nature Physics} \textbf{\bibinfo{volume}{5}},
  \bibinfo{pages}{805} (\bibinfo{year}{2009}).

\bibitem[{\citenamefont{Wouters and Carusotto}(2010)}]{Wouters:2010ee}
\bibinfo{author}{\bibfnamefont{M.}~\bibnamefont{Wouters}} \bibnamefont{and}
  \bibinfo{author}{\bibfnamefont{I.}~\bibnamefont{Carusotto}},
  \bibinfo{journal}{Physical Review Letters} \textbf{\bibinfo{volume}{105}},
  \bibinfo{pages}{020602} (\bibinfo{year}{2010}).

\bibitem[{\citenamefont{Avenel et~al.}(1997)\citenamefont{Avenel, Hakonen, and
  Varoquaux}}]{Avenel:1997io}
\bibinfo{author}{\bibfnamefont{O.}~\bibnamefont{Avenel}},
  \bibinfo{author}{\bibfnamefont{P.}~\bibnamefont{Hakonen}}, \bibnamefont{and}
  \bibinfo{author}{\bibfnamefont{E.}~\bibnamefont{Varoquaux}},
  \bibinfo{journal}{Physical Review Letters} \textbf{\bibinfo{volume}{78}},
  \bibinfo{pages}{3602} (\bibinfo{year}{1997}).

\bibitem[{\citenamefont{Packard and Vitale}(1992)}]{Packard:1992jz}
\bibinfo{author}{\bibfnamefont{R.}~\bibnamefont{Packard}} \bibnamefont{and}
  \bibinfo{author}{\bibfnamefont{S.}~\bibnamefont{Vitale}},
  \bibinfo{journal}{Physical Review B} \textbf{\bibinfo{volume}{46}},
  \bibinfo{pages}{3540} (\bibinfo{year}{1992}).

\bibitem[{\citenamefont{E.~Varoquaux}(2008)}]{Varoquaux:2008}
\bibinfo{author}{\bibfnamefont{G.~V.} \bibnamefont{E.~Varoquaux}},
  \bibinfo{journal}{UFN} \textbf{\bibinfo{volume}{178}}, \bibinfo{pages}{217}
  (\bibinfo{year}{2008}).

\bibitem[{\citenamefont{Dasbach et~al.}(2001)\citenamefont{Dasbach, Schwab,
  Bayer, and Forchel}}]{Dasbach:2001wt}
\bibinfo{author}{\bibfnamefont{G.}~\bibnamefont{Dasbach}},
  \bibinfo{author}{\bibfnamefont{M.}~\bibnamefont{Schwab}},
  \bibinfo{author}{\bibfnamefont{M.}~\bibnamefont{Bayer}}, \bibnamefont{and}
  \bibinfo{author}{\bibfnamefont{A.}~\bibnamefont{Forchel}},
  \bibinfo{journal}{Physical Review B} \textbf{\bibinfo{volume}{64}},
  \bibinfo{pages}{201309} (\bibinfo{year}{2001}).

\bibitem[{\citenamefont{Kaitouni et~al.}(2006)\citenamefont{Kaitouni,
  El~Da{\"\i}f, Baas, Richard, Paraiso, Lugan, Guillet, Morier-Genoud,
  Gani{\`e}re, Staehli et~al.}}]{Kaitouni:2006hr}
\bibinfo{author}{\bibfnamefont{R.~I.} \bibnamefont{Kaitouni}},
  \bibinfo{author}{\bibfnamefont{O.}~\bibnamefont{El~Da{\"\i}f}},
  \bibinfo{author}{\bibfnamefont{A.}~\bibnamefont{Baas}},
  \bibinfo{author}{\bibfnamefont{M.}~\bibnamefont{Richard}},
  \bibinfo{author}{\bibfnamefont{T.}~\bibnamefont{Paraiso}},
  \bibinfo{author}{\bibfnamefont{P.}~\bibnamefont{Lugan}},
  \bibinfo{author}{\bibfnamefont{T.}~\bibnamefont{Guillet}},
  \bibinfo{author}{\bibfnamefont{F.}~\bibnamefont{Morier-Genoud}},
  \bibinfo{author}{\bibfnamefont{J.}~\bibnamefont{Gani{\`e}re}},
  \bibinfo{author}{\bibfnamefont{J.}~\bibnamefont{Staehli}},
  \bibnamefont{et~al.}, \bibinfo{journal}{Physical Review B}
  \textbf{\bibinfo{volume}{74}}, \bibinfo{pages}{155311}
  (\bibinfo{year}{2006}).

\bibitem[{\citenamefont{Bajoni et~al.}(2008)\citenamefont{Bajoni, Senellart,
  Wertz, Sagnes, Miard, Lema{\^\i}tre, and Bloch}}]{Bajoni:2008ef}
\bibinfo{author}{\bibfnamefont{D.}~\bibnamefont{Bajoni}},
  \bibinfo{author}{\bibfnamefont{P.}~\bibnamefont{Senellart}},
  \bibinfo{author}{\bibfnamefont{E.}~\bibnamefont{Wertz}},
  \bibinfo{author}{\bibfnamefont{I.}~\bibnamefont{Sagnes}},
  \bibinfo{author}{\bibfnamefont{A.}~\bibnamefont{Miard}},
  \bibinfo{author}{\bibfnamefont{A.}~\bibnamefont{Lema{\^\i}tre}},
  \bibnamefont{and} \bibinfo{author}{\bibfnamefont{J.}~\bibnamefont{Bloch}},
  \bibinfo{journal}{Physical Review Letters} \textbf{\bibinfo{volume}{100}},
  \bibinfo{pages}{047401} (\bibinfo{year}{2008}).

\bibitem[{\citenamefont{Lai et~al.}(2007)\citenamefont{Lai, Kim, Utsunomiya,
  Roumpos, Deng, Fraser, Byrnes, Recher, Kumada, Fujisawa et~al.}}]{Lai:2007dd}
\bibinfo{author}{\bibfnamefont{C.~W.} \bibnamefont{Lai}},
  \bibinfo{author}{\bibfnamefont{N.~Y.} \bibnamefont{Kim}},
  \bibinfo{author}{\bibfnamefont{S.}~\bibnamefont{Utsunomiya}},
  \bibinfo{author}{\bibfnamefont{G.}~\bibnamefont{Roumpos}},
  \bibinfo{author}{\bibfnamefont{H.}~\bibnamefont{Deng}},
  \bibinfo{author}{\bibfnamefont{M.~D.} \bibnamefont{Fraser}},
  \bibinfo{author}{\bibfnamefont{T.}~\bibnamefont{Byrnes}},
  \bibinfo{author}{\bibfnamefont{P.}~\bibnamefont{Recher}},
  \bibinfo{author}{\bibfnamefont{N.}~\bibnamefont{Kumada}},
  \bibinfo{author}{\bibfnamefont{T.}~\bibnamefont{Fujisawa}},
  \bibnamefont{et~al.}, \bibinfo{journal}{Nature}
  \textbf{\bibinfo{volume}{450}}, \bibinfo{pages}{529} (\bibinfo{year}{2007}).

\bibitem[{\citenamefont{Amo et~al.}(2010)\citenamefont{Amo, Pigeon, Adrados,
  Houdre, Giacobino, Ciuti, and Bramati}}]{Amo:2010ff}
\bibinfo{author}{\bibfnamefont{A.}~\bibnamefont{Amo}},
  \bibinfo{author}{\bibfnamefont{S.}~\bibnamefont{Pigeon}},
  \bibinfo{author}{\bibfnamefont{C.}~\bibnamefont{Adrados}},
  \bibinfo{author}{\bibfnamefont{R.}~\bibnamefont{Houdre}},
  \bibinfo{author}{\bibfnamefont{E.}~\bibnamefont{Giacobino}},
  \bibinfo{author}{\bibfnamefont{C.}~\bibnamefont{Ciuti}}, \bibnamefont{and}
  \bibinfo{author}{\bibfnamefont{A.}~\bibnamefont{Bramati}},
  \bibinfo{journal}{Physical Review B} \textbf{\bibinfo{volume}{82}},
  \bibinfo{pages}{081301} (\bibinfo{year}{2010}).

\bibitem[{\citenamefont{Keeling and Berloff}(2008)}]{Keeling:2008hj}
\bibinfo{author}{\bibfnamefont{J.~M.~J.} \bibnamefont{Keeling}}
  \bibnamefont{and} \bibinfo{author}{\bibfnamefont{N.~G.}
  \bibnamefont{Berloff}}, \bibinfo{journal}{Physical Review Letters}
  \textbf{\bibinfo{volume}{100}}, \bibinfo{pages}{250401}
  (\bibinfo{year}{2008}).

\bibitem[{\citenamefont{Wouters and Carusotto}(2007)}]{wouters07:goldstone}
\bibinfo{author}{\bibfnamefont{M.}~\bibnamefont{Wouters}} \bibnamefont{and}
  \bibinfo{author}{\bibfnamefont{I.}~\bibnamefont{Carusotto}},
  \bibinfo{journal}{Physical Review A} \textbf{\bibinfo{volume}{76}},
  \bibinfo{pages}{043807} (\bibinfo{year}{2007}).

\bibitem[{\citenamefont{Wouters and Savona}(2010)}]{Wouters:2010vyxx}
\bibinfo{author}{\bibfnamefont{M.}~\bibnamefont{Wouters}} \bibnamefont{and}
  \bibinfo{author}{\bibfnamefont{V.}~\bibnamefont{Savona}}
  (\bibinfo{year}{2010}), \eprint{arXiv:1007.5453v1}.

\bibitem[{\citenamefont{Borgh et~al.}(2010)\citenamefont{Borgh, Keeling, and
  Berloff}}]{Borgh:2010jv}
\bibinfo{author}{\bibfnamefont{M.~O.} \bibnamefont{Borgh}},
  \bibinfo{author}{\bibfnamefont{J.~M.~J.} \bibnamefont{Keeling}},
  \bibnamefont{and} \bibinfo{author}{\bibfnamefont{N.~G.}
  \bibnamefont{Berloff}}, \bibinfo{journal}{Physical Review B}
  \textbf{\bibinfo{volume}{81}}, \bibinfo{pages}{235302}
  (\bibinfo{year}{2010}).

\bibitem[{\citenamefont{Balili et~al.}(2006)\citenamefont{Balili, Snoke,
  Pfeiffer, and West}}]{Balili:2006dm}
\bibinfo{author}{\bibfnamefont{R.~B.} \bibnamefont{Balili}},
  \bibinfo{author}{\bibfnamefont{D.~W.} \bibnamefont{Snoke}},
  \bibinfo{author}{\bibfnamefont{L.}~\bibnamefont{Pfeiffer}}, \bibnamefont{and}
  \bibinfo{author}{\bibfnamefont{K.}~\bibnamefont{West}},
  \bibinfo{journal}{Applied Physics Letters} \textbf{\bibinfo{volume}{88}},
  \bibinfo{pages}{031110} (\bibinfo{year}{2006}).

\bibitem[{\citenamefont{Lega and Fauve}(1997)}]{Lega:1997}
\bibinfo{author}{\bibfnamefont{J.}~\bibnamefont{Lega}} \bibnamefont{and}
  \bibinfo{author}{\bibfnamefont{S.}~\bibnamefont{Fauve}},
  \bibinfo{journal}{Physica D} \textbf{\bibinfo{volume}{102}},
  \bibinfo{pages}{234} (\bibinfo{year}{1997}).

\end{thebibliography}

\end{document}